\documentclass[twocolumn,english,aps,prb,superscriptaddress,natbib,bibnotes,amsmath,amssymb,floatfix,groupedaddress,footinbib]{revtex4-2}

\usepackage[colorlinks=true,citecolor=blue,linkcolor=magenta]{hyperref}

\usepackage[markup=nocolor, authormarkupposition=left]{changes} 

\usepackage{soul}
\usepackage[utf8]{inputenc}
\usepackage[english]{babel}
\usepackage{amsmath,amsfonts,amssymb}
\usepackage[T1]{fontenc}
\usepackage{url}

\usepackage{amsmath}
\usepackage{siunitx}
\usepackage[version=4]{mhchem}
\usepackage{amsfonts}
\usepackage{amssymb}

\usepackage{epstopdf}
\usepackage{graphicx}
\graphicspath{{./Figures/}}
\usepackage{gensymb}
\usepackage{threeparttable}

\usepackage{changes}
\DeclareUnicodeCharacter{2212}{-}
\DeclareUnicodeCharacter{2009}{\hspace{0.2em}}  
\begin{document}

\title{High-Efficiency Polarization-Diversity Grating Coupler with Multipolar Radiation Mode Enhancement}

\author{Wu Zhou,\textsuperscript{1} Kaihang Lu,\textsuperscript{1} Shijie Kang,\textsuperscript{2} Xiaoxiao Wu,\textsuperscript{2} and Yeyu Tong\textsuperscript{1}}
\email{wzhou832@connect.hkust-gz.edu.cn}
\email{yeyutong@hkust-gz.edu.cn}
\affiliation{\textsuperscript{1}Microelectronic Thrust, The Hong Kong University of Science and Technology (Guangzhou), Guangzhou, Guangdong, 511453, China.}
\affiliation{\textsuperscript{2}Advanced Materials Thrust, The Hong Kong University of Science and Technology (Guangzhou), Guangzhou, Guangdong, 511453, China.}

\maketitle

\noindent\textbf{\noindent Two-dimensional (2D) diffraction gratings offer a polarization-independent coupling solution between the planar photonic chips and optical fibers, with advantages including placement flexibility, ease of fabrication, and tolerance to alignment errors. In this work, we first proposed and experimentally demonstrated a highly efficient 2D grating coupler enabled by exciting multipolar resonances through bi-level dielectric structures. A 70-nm shallow-etched hole array and a 160-nm-thick deposited polycrystalline silicon tooth array are employed in our proposed 2D grating coupler. Strong optical field confinement and enhanced radiation directionality can thus be attained through the use of 193-nm deep-ultraviolet (DUV) lithography, which is readily accessible from commercial silicon photonics foundries. The measured experimental peak coupling efficiency is -2.54 dB with a minimum feature size of 180 nm. Our design exhibits a 3-dB bandwidth of around 23.4 nm with good positioning tolerance for optical fibers. Due to the benefits of perfectly vertical coupling, the measured polarization-dependent loss in our experiments is below 0.3 dB within the 3-dB working bandwidth. Our proposed 2D grating structure and design method can also be applied to other integrated optics platforms, enabling an efficient and polarization-diversity coupling between optical fibers and photonic chips while reducing requirements on feature size.}

\section*{Introduction} 

Silicon photonics platform has experienced significant growth over the last decade, with high-bandwidth optical transceivers now being mass-produced and transitioning into on-board and in-package optical interconnects \cite{xiang2021perspective,wade2021error,shi2022silicon}. Various emerging applications, such as programmable photonic processors \cite{bogaerts2020programmable,lu2024empowering} and prototypes for optical sensing and imaging \cite{sun2013large, rogers2021universal} have been successfully demonstrated. The appeal of silicon photonics lies in its ability to leverage the well-established complementary metal-oxide-semiconductor (CMOS) compatible fabrication processes. Additionally, the high refractive index contrast enables the strong confinement of optical field, resulting in the reduction of the form factor of optical components and circuits. However, this has also caused a significant challenge in achieving optimal fiber-chip coupling interfaces due to the huge difference in mode size between optical fibers and silicon waveguides. 

Both in-plane and out-of-plane coupling strategies have been extensively researched over the years \cite{marchetti2019coupling, kopp2010silicon}. While edge couplers exhibit low polarization dependence and high coupling efficiency, they typically require high-quality facets on the chip sides. Out-of-plane coupling, utilizing diffractive gratings, has been widely adopted in wafer-scale testing due to its advantages of simple back-end processing, position flexibility and low alignment accuracy requirements \cite{kopp2010silicon, benedikovic2017shaped}. Nevertheless, one-dimensional (1D) grating coupler typically manifests a strong polarization dependence and work optimally only for certain linear polarization state \cite{benedikovic2015high}. Subwavelength gratings can support both transverse electric (TE) and transverse magnetic (TM) modes, allowing for polarization-insensitive fiber-chip coupling \cite{cheng2014experimental, mak2018multi, wang2015apodized}. However, additional polarization splitters and rotators are necessary to convert the TM mode into the TE mode, as most photonic integrated components for optical interconnects are designed for the TE mode \cite{marchetti2019coupling}. Two-dimensional (2D) grating couplers provide an alternative solution for polarization-insensitive coupling \cite{taillaert2003compact}. They can be viewed as a superposition of two 1D grating couplers placed perpendicular to each other, thus can convert any arbitrary polarized light from the optical fiber into two orthogonal TE modes on a photonic chip. 

Nonetheless, the primary challenge with 2D grating couplers is achieving a high coupling efficiency while accommodating the 193 nm deep-ultraviolet (DUV) lithography that is currently provided by the majority of commercial silicon photonics foundries \cite{rahim2019open}. While it has been shown that coupling loss can be effectively reduced to around -2.5 dB per interface by either increasing the thickness of the crystalline silicon layer to 340 nm \cite{zhang2021high, zhang2020two}, or by employing blazed subwavelength structures with a minimum feature size of 40 nm \cite{watanabe20192}, these approaches do not provide a direct solution for the prevalent 220-nm silicon-on-insulator (SOI) wafers that are based on the 130 nm or 180 nm CMOS node toolset \cite{rahim2019open}. To address this issue, the double-layer structure previously applied for 1D grating couplers \cite{sacher2014wide, sacher2018monolithically, tong2018efficient,vitali2023high,zhou2022photolithography} may be utilized, which involves depositing additional layers of polycrystalline silicon (poly-Si) or silicon nitride ($\mathrm{Si_{3}N_{4}}$) on top of the SOI platform. These layers can break the vertical symmetry of the structure to facilitate an improved directionality, defined here as the ratio of upward-diffracted light energy to the total radiated optical power. The deposition of additional poly-Si or $\mathrm{Si_{3}N_{4}}$ has become a de facto standard in commercial silicon photonic foundries. However, it is difficult to implement a bi-level structure for 2D grating coupler while also adhering to the fabrication constraints, which has been investigated in simulations \cite{carroll2014optimizing,hammond2022multi} but never yet been experimentally realized, to the best of our knowledge.

In this work, we present the first experimental demonstration of a highly efficient and polarization-diversity 2D grating coupler fabricated on a 220-nm-thick SOI platform using 193-nm DUV lithography for standard single-mode fibers (SMFs). By employing a carefully engineered structure to excite multipolar radiation modes, the grating achieves enhanced directionality and superior fiber-to-chip coupling performance. The demonstrated fiber-to-chip coupling efficiency reaches -2.54 dB (56\%) with a 3-dB bandwidth of 23.4 nm. Measurements indicate a polarization-dependent loss (PDL) of less than 0.3 dB, attributed to the optimized vertical coupling configuration. The proposed 2D grating comprises 70-nm shallowly etched holes in the 220-nm SOI layer, with 160-nm-thick, aligned poly-Si teeth deposited on top. Detailed design considerations and simulation results are discussed in Section II, followed by fabrication details and experimental performance validation in Section III.

\section*{Design and Simulation}

\begin{figure}[b]
  \begin{center}
  \includegraphics[width=1.0\linewidth]{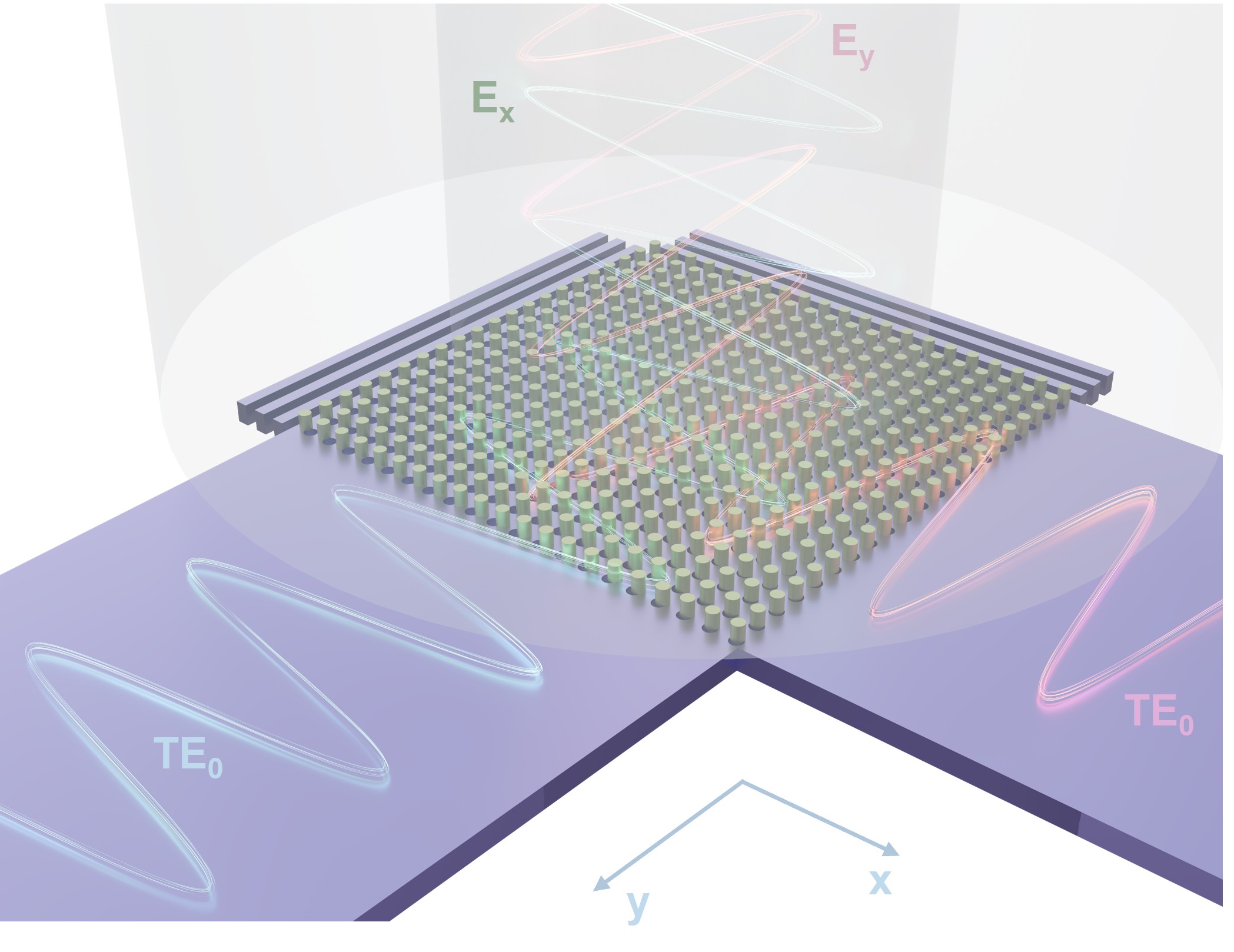}\\
  \caption{Schematic diagram of the perfectly vertical two-dimensional (2D) grating coupler for polarization-independent coupling with the optical fiber.}\label{gating_diagram}
  \end{center}
\end{figure}

\begin{figure*}[t]
  \centering
  \includegraphics[width=1\linewidth]{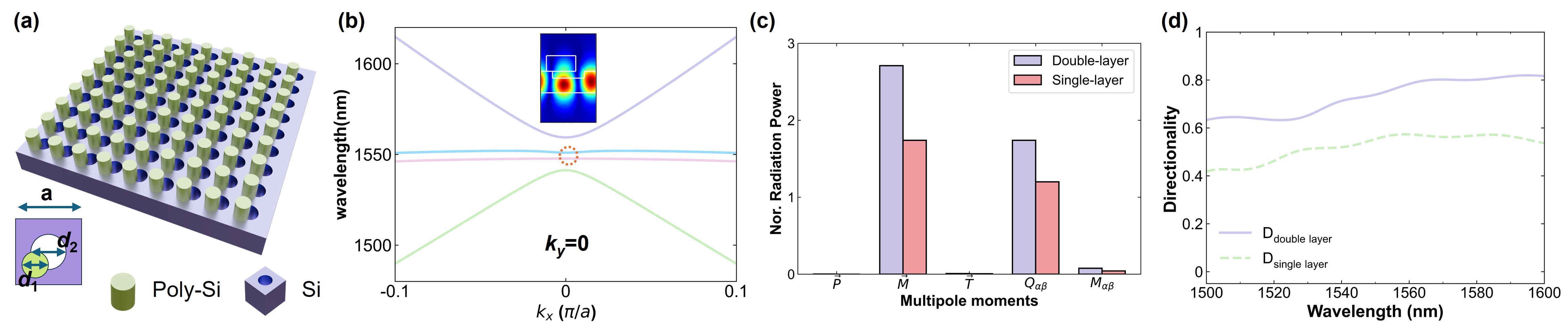}\\
  \caption{(a) Schematic illustration of the 2D photonic crystal structure, consisting of a silicon tooth array and a shallow-etched silicon hole array. (b) Band structure near the Gamma point for the structure shown in (a), with \(d_1 = 300 \, \mathrm{nm}\), \(d_2 = 325 \, \mathrm{nm}\), and \(a = 536 \, \mathrm{nm}\). The Gamma point (\(\Gamma\) point) is located at the center of the first Brillouin zone, where the wave vector \(\mathbf{k} = 0\). (c) Decomposed radiation power in terms of multipoles corresponding to the band structure shown in (b) near the 1550nm. (d) Comparison of the directivity of the designed double-layer structure, calculated using 3D FDTD simulations, with that of the traditional single-layer structure. Both structures have an identical 70-nm shallow-etched layer.}
  \label{band_structure}
\end{figure*}

Fig.~\ref{gating_diagram} illustrates the conceptual diagram of the 2D grating coupler interfaced with a SMF in perfectly vertical coupling configuration. The orthogonal polarizations, E\textsubscript{x} and E\textsubscript{y} in the SMF are coupled into the fundamental TE modes of two perpendicular silicon waveguides on a photonic chip. A low PDL can be maintained due to the adopted perfectly vertical coupling. To ensure uniform coupling performance for both polarizations, the grating is symmetrically designed along its diagonal axis. The device is designed to be fabricated on a 220-nm-thick SOI platform with a buried-oxide thickness of 2 $\mu$m. The 70-nm shallow etch process creates the low-refractive-index region within the grating structure. The coupling efficiency (CE) of grating couplers is primarily influenced by their directionality and modal overlap with the fiber mode. To enhance upward diffraction, alternatively bottom mirrors can be deposited beneath the grating region~\cite{luo2018low}. In our proposed design, we employ a two-dimensional (2D) photonic crystal structure to excite multipolar resonances, enabling stronger field confinement and enhanced radiation directionality. Specifically, we investigate a double-layer 2D photonic crystal structure, comprising a shallow etched silicon hole array and a silicon tooth array, as shown in Fig.~\ref{band_structure}. In this design, a poly-Si layer is utilized as the silicon tooth array. This approach eliminates the need for additional back-end-of-line (BEOL) processing steps, thereby simplifying fabrication. By carefully designing periodic structures, these multipolar radiation modes can be enhanced and precisely controlled. These result in significantly improved field confinement and radiation directionality, paving the way for higher efficiency in the proposed 2D grating coupler.

The structure shown in Fig.~\ref{band_structure}(a) is a photonic crystal array consisting of cylindrical poly-Si teeth with diameter \(d_1\), height \(h_1\), and periodicity \(a\) and a shallow etched silicon hole array with diameter \(d_2\), height \(h_2\), and periodicity \(a\). To identify the eigenmodes relevant to multipolar radiation in the C-band, a parameter sweep was performed to compute the photonic band structure near the Gamma point. As shown in Fig.~\ref{band_structure}(b), for a diameter of \(d_1 = 300 \, \mathrm{nm}\), \(d_2 = 325 \, \mathrm{nm}\), height of \(h_1 = 160 \, \mathrm{nm}\), \(h_2 = 70 \, \mathrm{nm}\) and periodicity \(a = 536 \, \mathrm{nm}\), the band structure reveals several quasi-degenerate eigenstates near the C-band. Among them, four eigenmodes, located at eigenfrequencies within the C-band near the Gamma-point, correspond to dielectric-based resonant modes. The zoom-in image of Fig.~\ref{band_structure}(b) shows the electric field distribution of the eigenmode at the Gamma-point for a wavelength of 1550 nm, exhibiting strongly localized electromagnetic fields. This is a characteristic signature of multipolar radiation. To further confirm these dielectric-based resonant modes, the radiated powers were decomposed into contributions from various multipole moments using general multipole scattering theory~\cite{li2015toroidal}:

\begin{gather*}
\quad \vec{P} = \frac{1}{i\omega} \int \vec{J} \, d^3r,
\quad \vec{M} = \frac{1}{2c} \int \left( \vec{r} \times \vec{J} \right) d^3r, \\[0em]
\quad \vec{T} = \frac{1}{10c} \int \left[ \left( \vec{r} \cdot \vec{J} \right) \vec{r} - 2r^2 \vec{J} \right] d^3r, \\[0em]
\end{gather*}

\begin{gather*}
\quad Q_{\alpha\beta} = \frac{1}{i2\omega} \int \left[ r_\alpha J_\beta + r_\beta J_\alpha - \frac{2}{3} \left( \vec{r} \cdot \vec{J} \right) \delta_{\alpha\beta} \right] d^3r, \\[0em]
\quad M_{\alpha\beta} = \frac{1}{3c} \int \left[ \left( \vec{r} \times \vec{J} \right)_\alpha r_\beta + \left( \vec{r} \times \vec{J} \right)_\beta r_\alpha \right] d^3r.
\end{gather*}

\noindent \(c\) represents the speed of light in vacuum, and \( \vec{r} \) is the position vector in Cartesian coordinates \( (x, y, z) \), with \( \alpha, \beta \in \{x, y, z\} \). The quantities \( \vec{P} \), \( \vec{M} \), \( \vec{T} \), \( Q_{\alpha\beta} \), and \( M_{\alpha\beta} \) correspond to the electric dipole moment, magnetic dipole moment, toroidal dipole moment, electric quadrupole moment, and magnetic quadrupole moment, respectively. The far-field radiated power, decomposed into these multipole moments, is then expressed as:

\begin{gather*}
I_P = 2 \omega^4 \frac{|\vec{P}|^2}{3c^3}, \quad
I_M = 2 \omega^4 \frac{|\vec{M}|^2}{3c^3}, \quad
I_T = 2 \omega^6 \frac{|\vec{T}|^2}{3c^5}, \\
I_Q^e = \omega^6 \frac{\sum |Q_{\alpha\beta}|^2}{5c^5}, \quad \text{and} \quad
I_Q^m = \omega^6 \frac{\sum |M_{\alpha\beta}|^2}{40c^5}.
\end{gather*}

\begin{figure}[t]
  \begin{center}
  \includegraphics[width=\linewidth]{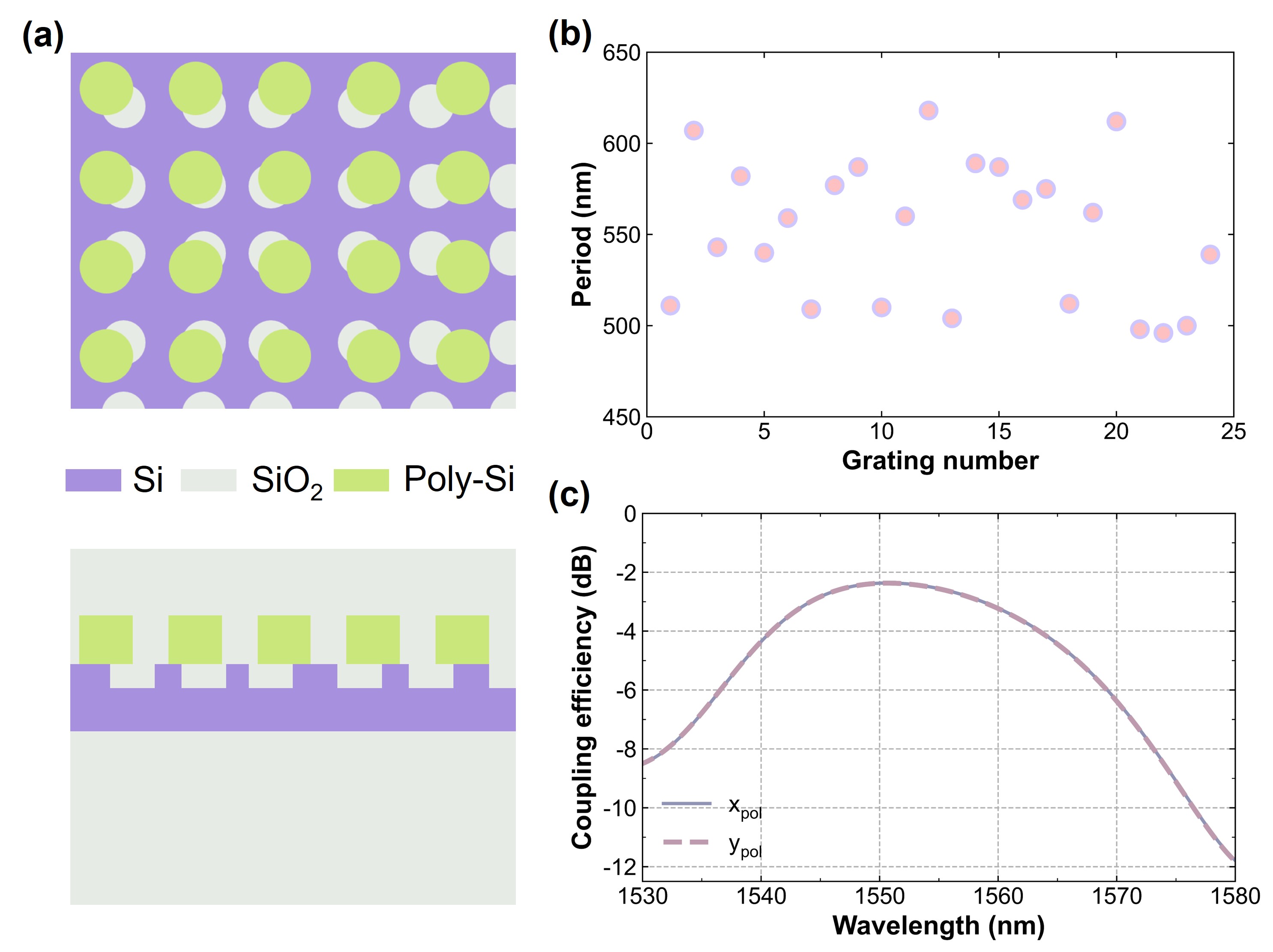}\\
  \caption{(a) Top view and cross-sectional diagram of the grating coupler. (b) Detailed distribution of each period within the designed grating coupler. (c) Coupling efficiency of the designed grating coupler obtained from 3D FDTD simulations.}\label{design and simulation}
  \end{center}
\end{figure}

Fig.~\ref{band_structure}(c) shows the decomposed powers of various multipole moments. The results indicate that, for the proposed photonic crystal array, the powers of the magnetic dipole and electric quadrupole moments are approximately one order of magnitude higher than those of other multipoles. This structure effectively excites specific multipolar radiation modes, with dominant contributions arising from the magnetic dipole and electric quadrupole moments. Fig.~\ref{band_structure}(d) illustrates the directionality obtained from 3D-FDTD simulations of the 2D photonic crystal structure. At a wavelength of 1550 nm, the directionality is approximately 75\%, representing a 20\% improvement compared to conventional single-layer structures. However, due to the insufficient overlap integral between the mode field and the fundamental mode of a single-mode fiber, the coupling efficiency of the designed 2D photonic crystal grating coupler remains relatively low.

\begin{figure*}[t]
  \begin{center}
  \includegraphics[width=\linewidth]{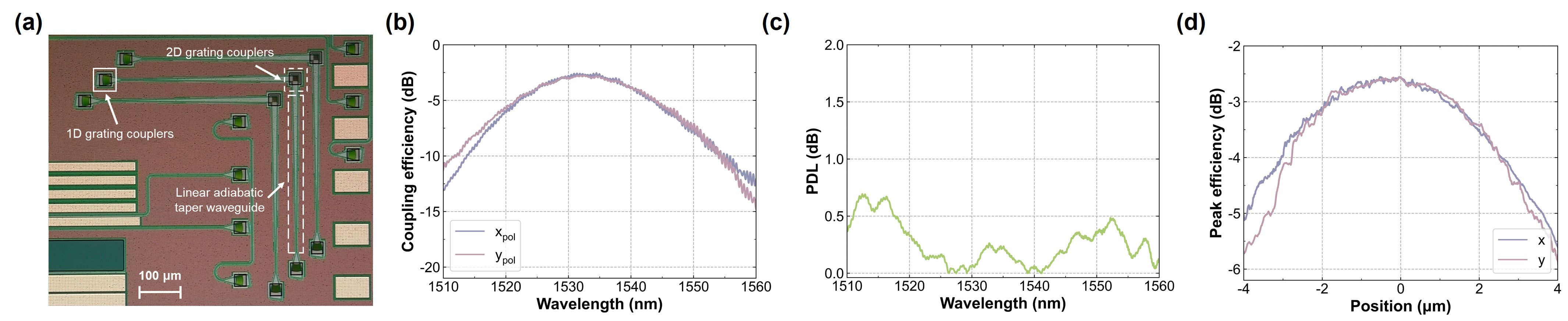}\\
  \caption{(a) Microscopic image of the fabricated photonic integrated circuits including the designed 2D grating coupler, 360-micron long linear adiabatic taper waveguide for spot size conversion, and 1D input grating coupler. (b) Measurement of the coupling efficiency for both polarizations of the 2D grating coupler. (c) Measurement of the Po  larization Dependent Loss (PDL) of the 2D grating coupler. (d) Measurement of the peak coupling efficiency sensitivity to the fiber position for the 2D grating coupler.}\label{fab and exp}
  \end{center}
\end{figure*}

To further enhance the coupling efficiency and suppress second-order Bragg reflections into the waveguides—caused by the perfect vertical coupling configuration \cite{chen2010two}—apodization is applied to the shallow-etched silicon hole array. Fig.~\ref{design and simulation}(a) illustrates the detailed structure of the double-layer 2D grating used for fabrication. The design features 70-nm-deep circular holes etched into a 220-nm-thick silicon layer, with a 160-nm-thick cylindrical poly-Si tooth array deposited on top. The poly-Si teeth are uniformly spaced with a consistent diameter, while the period of the circular holes is apodized to improve coupling performance. To further enhance directionality and modal overlap, a silicon reflector is integrated at the end of the grating. The key structural parameters were optimized, including the periodicity of the nanoholes (\(p_i\)) and the width and spacing of the end reflectors, as shown in Fig.~\ref{design and simulation}(b). Other parameters, such as the hole diameter (\(d_{\text{hole}}\)), teeth diameter (\(d_{\text{teeth}}\)), and teeth periodicity (\(p_{\text{teeth}}\)), were kept constant with previously calculated values: \(d_{\text{hole}} = 325 \, \mathrm{nm}\), \(d_{\text{teeth}} = 300 \, \mathrm{nm}\), and \(p_{\text{teeth}} = 536 \, \mathrm{nm}\). Design optimization was carried out using a genetic algorithm combined with finite-difference time-domain (FDTD) simulations \cite{Tong2020Efficient}. To address the high computational cost of 3D FDTD simulations—required for modeling both optical fibers and silicon waveguide gratings—the optimization process was simplified by incorporating Effective Medium Theory (EMT) for 2D FDTD simulations \cite{chen2010two}. This approach allowed for a fast and efficient evaluation of the Figure of Merit (FOM) during optimization, significantly reducing computational expenses while enabling a thorough search for the optimal design parameters.

Fig.~\ref{design and simulation}(c) illustrates the length of each grating period for the 70-nm shallowly etched holes in the final design. The end reflector has a width of 365 nm and a spacing of 360 nm. The proposed 2D grating consists of 24 periods, resulting in a grating area width of approximately 13.2 µm, which ensures a good match between the diffracted optical field size and the mode field diameter of the SMF. To accurately evaluate the performance of the optimized device, a 3D FDTD simulation was conducted to assess the coupling efficiency spectra for the two orthogonal polarizations. Fig.~\ref{design and simulation}(d) shows the spectrum for the two orthogonal polarizations when the optical signal is coupled from the optimized 2D grating coupler into an SMF. The predicted peak coupling efficiency is -2.37 dB (58\%) at a wavelength of 1550 nm, with a 3-dB bandwidth of 30 nm. The symmetrical design ensures consistent performance for both polarizations. 

\section{Fabrication and Measurement}

The proposed double-layer 2D grating coupler was fabricated using a multi-project wafer (MPW) run provided by imec. Fig.~\ref{fab and exp}(a) displays a microscopic image of the fabricated 2D grating coupler, which is connected to two 1D grating couplers with a coupling angle of 10 degree. A 360-$\mu$\textit{m} long linear adiabatic taper waveguide is utilized as the spot size converter to match the mode field size in the SMF.

The coupling efficiency of the 2D grating coupler is characterized by measuring the fiber-chip-fiber transmission spectrum with a tunable laser source (Santec TSL-570) and an optical power meter (Santec MPM-210H and MPM-215). A three-paddle manual polarization controller (Thorlabs FPC560) is used to alter the polarization state of the optical signal in the SMF. After normalizing out the insertion loss of the 1D grating coupler, Fig.~\ref{fab and exp}(b) shows the coupling efficiency spectra for the two orthogonal polarizations E\textsubscript{x} and E\textsubscript{y}. A peak coupling efficiency of -2.54 dB at 1535 nm can be obtained for the x polarization. Compared to the simulation results, a small wavelength shift in the peak wavelength is observed. This deviation may arise from the fabrication uncertainties. The measured 3-dB bandwidth of the 2D grating coupler is 23.4 nm. Due to the symmetrical design for the two orthogonal polarizations, the y-polarization shows a similar peak coupling efficiency of -2.7 dB at 1531 nm. Such small discrepancies between the two polarizations are mainly due to the imperfections in the measurement alignment and fabrication process. 

\begin{table*}[t]
\caption{Comparison of the Polarization-Diversity 2D Grating Couplers on 220-nm SOI$^\textit{a}$}
\label{table 1}
\renewcommand{\arraystretch}{2}
\centering
\begin{tabular}{cccccccccc}
\hline
Ref. & Year & WB & CA [$^\circ$] & CE [dB] & CE [{\%}] & BW [nm] & PDL [dB] & Platform & MFS [nm]\\
\hline
\citenum{van2009focusing}  & 2009 & C & 10 & -5.6 & 27\% & - & \textgreater 0.4  & 220 nm SOI & 180\\
\citenum{chen2010two}  & 2010 & C & 0 & -5.2 & 30\% & - & \textless 0.4 & 220 nm SOI & 195\\
\citenum{verslegers2014design}  & 2014 & S & 8 & -2 & 63\% & 26.6$^{\ast}$ & - & Double SOI & -\\
\citenum{wu2016cmos}  & 2016 & O & 6 & -3.3 & 47\% & 43$^{\star}$ &  \textless 1.2 & 220 nm SOI & 180\\
\citenum{zouhust2016} & 2016 & C & 14 & -5 & 32\% & - &  \textless 0.25 & 220 nm SOI & $\approx 200$\\
\citenum{lacava2016design}  & 2016 & C & 14 & -3.75 & 42\% & 43$^{\star}$ & -  & 220 nm SOI & 272\\
\citenum{sobu2018si}  & 2018 & O & 10 & -3.9 & 41\% & - & \textless 0.3  & 200 nm SOI & -\\
\citenum{watanabe20192}  & 2019 & C & 0 & -2.6 & 55\% & 27$^{\star}$ & 3  & 220 nm SOI & 40\\
\citenum{xue2019two}  & 2019 & C & 12 & -4.2 & 38\% & - & \textless 0.2  & 220 nm SOI & 310\\
\citenum{ruiz2021compact}  & 2021 & C & 10 & -4.3 & 37\% & 48$^{\star}$ & \textless 0.54  & 250 nm SOI & 187\\
\citenum{zhang2021high}  & 2021 & C & 0 & -3.1 & 49\% & 42$^{\ast}$ & \textless 0.2  & 220 nm SOI & 180\\
\citenum{hammond2022multi} & 2022 & C & 0 & -7.0 & 20\% & 89$^{\star}$  & 1.9  & GF 45SPCLO + PolySi & -\\
This work & 2024 & C & 0 & -2.54 & 56\% & 23.4$^{\star}$ & \textless 0.3  & 220 nm SOI + PolySi & 180\\
\hline
\end{tabular}
 \label{tab:shape-functions}

$^\textit{a}$WB: wavelength band; C:1530-1565 nm; S:1460-1530 nm; O:1260-1360 nm; CA: coupling angle; \\CE: coupling efficiency; MFS: minimum feature size; BW: bandwidth, $^{\ast}$: 1-dB bandwidth, $^{\star}$: 3-dB bandwidth.
\end{table*}

The PDL is also characterized by the ratio between the maximum transmission($T_{max}$) and the minimum transmission ($T_{min}$) when varying fiber polarization across all possible states \cite{VanLaere2009Silicon}. In the experiment, a randomly polarized optical mode from the fiber was injected into the same 2D grating coupler, and the total transmission of the two output ports was measured. Fig.~\ref{fab and exp}(c) presents the spectrum of the observed PDL. Overall, the PDL is less than 0.3 dB within the 3-dB bandwidth of the 2D grating, and remains below 0.75 dB from 1510 nm to 1560 nm wavelength range. This small PDL can be attributed to the asymmetric coupling efficiency between the x-port and y-port, as well as potential measurement errors and fabrication defects given that the theoretical PDL for our design is expected to be zero.

Finally, we investigated the sensitivity of coupling efficiency versus the fiber misalignment for our 2D grating coupler. Light is sent from an SMF into the 2D grating coupler, and the optical polarization is optimized for E\textsubscript{x} or E\textsubscript{y}. After maximizing the coupled power by adjusting the position of the SMF, the coupling efficiency is measured against various fiber positions by a piezoelectric stage. Fig. \ref{fab and exp}(d) shows the variation of peak coupling efficiency with respect to fiber position. It can be observed that within a 2-µm positional deviation, the peak coupling efficiency remains within -3.5 dB. This indicates that the demonstrated 2D silicon grating coupler has a low alignment accuracy requirement, suggesting its potential of using a low-cost packaging process for polarization-diversity chip-fiber coupling.

Compared to the performance of 2D grating couplers reported in previous works on 220-nm SOI, as summarized in Table~\ref{table 1}, our proposed 2D grating coupler achieves the highest experimental peak coupling efficiency using a minimum feature size of 180 nm. Additionally, the measured polarization-dependent loss (PDL) is below 0.3 dB within its 3-dB bandwidth. To further enhance the coupling efficiency, future optimizations could focus on the poly-Si layer thickness \cite{carroll2014optimizing}. Furthermore, utilizing focusing 2D gratings could enable further reduction of the device footprint, eliminating the need for long linear adiabatic spot-size converters \cite{VanLaere2009Silicon}.

\section{Conclusion}

In summary, this work addresses the challenge of a highly efficient and polarization-diversity fiber-chip coupling by relying on the multipolar radiation modes in a double-layer 2D grating coupler. The demonstrated fiber-to-chip coupling efficiency can reach -2.54 dB (56\%) with a minimum feature size of 180 nm. The measured experimental 3-dB bandwidth is 23.4 nm with a polarization-dependence loss \textless 0.3 dB. Our proposed design method of leveraging multipolar radiation modes can also be applied to other integrated optics platforms, such as $\mathrm{Si_{3}N_{4}}$ on SOI or thin-film lithium niobate, to realize efficient polarization-diverse interfaces between optical fibers and photonic integrated circuits.

\vspace{0.2cm}

\begin{acknowledgments}
The authors acknowledge imec for photonic integrated circuits fabrication. This work was funded by the National Natural Science Foundation of China (No. 62305277, 12304348), Natural Science Foundation of Guangdong Province (No.2024A1515012438), Department of Education of Guangdong Province (No.2023KTSCX167), Guangzhou Association for Science and Technology (No. QT2024-011), and Guangzhou Municipal Science and Technology Project (No.2023A03J0013).
\end{acknowledgments}

\section*{Conflict of Interest}
The authors have no conflicts to disclose.

\section*{Author Contributions}
\textbf{Wu Zhou}: Conceptualization(equal), Theoretical calculation(equal), Experimental validation(lead), Writing – original draft(equal); \textbf{Kaihang Lu}: Theoretical calculation(supporting)), Experimental validation(supporting)), Writing – review \& editing(supporting); \textbf{Shijie Kang}: Theoretical calculation(equal)), Writing – review \& editing(supporting); \textbf{Xiaoxiao Wu}: Theoretical calculation(equal)), Writing – review \& editing(supporting), Supervision(supporting), Funding acquisition(supporting); \textbf{Yeyu Tong}: Conceptualization(equal), Theoretical calculation(equal), Experimental validation(supporting), Writing – original draft(equal), Funding acquisition(lead), Supervision(lead).

\section*{Data Availability Statement}
The data that support the findings of this study are available from the corresponding author upon reasonable request.

\bibliography{ref}
\end{document}